\documentclass[aps,prl,twocolumn,superscriptaddress,showpacs]{revtex4}
\def\be{\begin{equation}}
\def\ee{\end{equation}}
\def\bea{\begin{eqnarray}}          
\def\eea{\end{eqnarray}}
\def\bi{\begin{itemize}}
\def\ei{\end{itemize}}

\usepackage{graphicx}

\begin{document}

\title{ Winding up by a quench: 
        Insulator to superfluid phase transition in a ring of BECs }

\author{Jacek Dziarmaga} 
\affiliation{Institute of Physics and Centre for Complex Systems Research, 
             Jagiellonian University, Reymonta 4, 30-059 Krak\'ow, Poland}
\affiliation{Theory Division, Los Alamos National Laboratory, Los Alamos, NM 87545, USA}

\author{Jakub Meisner}
\affiliation{Institute of Physics and Centre for Complex Systems Research, 
             Jagiellonian University, Reymonta 4, 30-059 Krak\'ow, Poland}
\affiliation{Theory Division, Los Alamos National Laboratory, Los Alamos, NM 87545, USA}

\author{Wojciech H. Zurek}
\affiliation{Theory Division, Los Alamos National Laboratory, Los Alamos, NM 87545, USA}

\date{ April 22, 2008 }

\begin{abstract}
We study phase transition from the Mott insulator to superfluid in a periodic optical lattice. 
Kibble-Zurek mechanism predicts buildup of winding number through random walk of BEC phases, 
with the step size scaling as a the third  root of transition rate. We confirm this and demonstrate that 
this scaling accounts for the net winding number after the transition. 
\end{abstract}

\maketitle


{\bf Introduction. ---} 
In a second order phase transition, the critical point is characterized by divergences 
in the correlation length and in the relaxation time. This critical slowing down implies 
that no matter how slowly a system is driven through the transition its evolution cannot 
be adiabatic close to the critical point \cite{Z}. As a result, the state after the 
transition is not perfectly ordered: it is a mosaic of domains whose size depends on 
the rate of the transition. This scenario was first described in the cosmological 
setting by Kibble \cite{K} who appealed to relativistic causality to set an upper bound on 
domain size. The dynamical mechanism that determines domain size in second order phase transitions was 
proposed by one of us \cite{Z}. It is based on the universality of critical slowing down, 
and predicts that average size of the ordered domains $\hat\xi$ scales with the 
transition time $\tau_Q$ as $\hat\xi\sim\tau_Q^w$, where $w$ is a combination of 
critical exponents. This Kibble-Zurek mechanism (KZM) for second order thermodynamic 
phase transitions was confirmed by numerical simulations \cite{KZnum} and 
tested by experiments in liquid crystals \cite{LC}, superfluid helium 3 \cite{He3}, 
both high-$T_c$ \cite{highTc} and low-$T_c$ \cite{lowTc} superconductors, and even 
in non-equilibrium systems \cite{ne}. With the exception of superfluid $^4$He -- where 
the situation remains unclear \cite{He4a}, experimental results are consistent 
with KZM (see \cite{Kibble07} for a review). Spontaneous appearance of vorticity during 
Bose-Einstein condensation driven by evaporative cooling was recently reported 
\cite{Anderson}. This confirms KZM predictions \cite{BEC}, and is further elucidated by 
numerical studies of BEC formation \cite{Bradley}.

Our goal is to study dynamics of a quantum phase transition in a simple yet non-trivial 
example that can be implemented experimentally. Quantum phase transitions we consider 
differ qualitatively from finite temperature 
transitions. Most importantly, evolution is unitary, so there is no damping, and no 
thermal fluctuations to initiate symmetry breaking. Recent work on the dynamics of 
quantum phase transitions is mostly theoretical, \cite{3sites,Bodzio1,KZIsing, 
Dziarmaga2005,Polkovnikov,Cucchietti,Levitov,Cincio,Fubini,random}, but there is one 
possible exception: Ref. \cite{ferro} on the transition in a spin-$1$ BEC. Generic 
outcome of that experiment is a mosaic of ferromagnetic domains whose origin was 
attributed to a sudden quench limit of KZM. This explanation is supported by theory 
\cite{Lamacraft}.


{\bf Model. ---}
Bose-Hubbard model is a paradigmatic example of a non-integrable quantum critical 
system. It describes cold bosonic atoms in an optical lattice \cite{Greiner}. 
In dimensionless variables, its Hamiltonian reads
\begin{equation}
\label{H}
H = -J \sum_{s=1}^N \left( a_{s+1}^\dag a_s + {\rm h. c.} \right)
  + \frac{1}{2n} \sum_{i=1}^N a_s^\dag a_s^\dag a_s a_s ~.
\end{equation}
Here $N$ is the number of lattice sites and $n$ is an average number of atoms per site. 
This model with periodic boundary conditions (which we assume) should be directly 
experimentally accessible in a ring-shaped optical lattice \cite{amico}. For an integer 
$n$, the transition from the Mott insulator (small $J$) to the superfluid phase 
(large $J$) is located at $J_c\simeq n^{-2}$ \cite{KT}. 

We drive the system through its critical point by a linear quench with a quench 
timescale $\tau_Q$:
\be
J(t)~=~{t}/{\tau_Q}
\label{Jt}
\ee
In an experiment one can increase Josephson coupling $J$ by turning off the optical 
lattice potential as in \cite{Greiner}. The initial state is the Mott insulator ground 
state at $J=0$, 
\be
|n,n,n,\dots,n\rangle~,
\label{Mott}
\ee
with the same atom number at each site. We assume $n\gg 1$: This large density limit 
is accessible experimentally.


{\bf Numerical approach. ---}
We replace annihilation operators $a_s$ by complex field $\phi_s$,
$
a_s\approx\sqrt{n}~\phi_s~,
$ 
which is normalized,
$
\sum_{s=1}^N |\phi_s|^2=N~,
$
and evolves with the time-dependent Gross-Pitaevskii equation 
\be
i\frac{d\phi_s}{dt}=
-J\left(\phi_{s+1}+\phi_{s-1}-2\phi_s\right)+
|\phi_s|^2\phi_s~.
\label{GPE}
\ee
These approximations are accurate for $n\to\infty$, when the critical point
$J_c\simeq n^{-2}\to 0$.

In the truncated Wigner method we employ quantum expectation values are given 
by the averages over stochastic realisations of the field $\phi_s(t)$ 
\cite{Kazik,TW,Bradley}. For example, the correlation function becomes
\be
C_R ~=~
\frac{\langle a_s^\dag a_{s+R}\rangle}{n}~\approx~
\overline{\phi_s^*\phi_{s+R}}~.
\label{CR}
\ee
Here $\langle..\rangle$ means quantum expectation value while the overline is an 
average over realizations. All realizations of $\phi_s(t)$ evolve with the same 
deterministic Gross-Pitaevskii equation (\ref{GPE}), but they start from different 
random initial conditions which come from a probability distribution depending on 
an initial quantum state. The initial Mott state (\ref{Mott}) corresponds to initial 
fields
\be
\phi_s(0)~=~e^{i\theta_s}~
\label{randomphases}
\ee
with independent random phases $\theta_s\in[0,2\pi)$: The Mott state has 
the same number of $n$ particles at each site (i.e., $|\phi_s(0)|=1$), and, hence, 
indeterminate phases, that translate into random $\theta_s$.


{\bf Kibble-Zurek mechanism. ---}
In an optical lattice with BEC pools that become gradually connected with Josephson 
couplings in accord with Eq.~(\ref{Jt}) it is natural to rephrase KZM: Rather than 
seek distance $\hat\xi$ over which phase remains more or less the same we compute 
size $\Delta\theta_s$ of a typical phase step between neighboring sites. One could 
use it to deduce the size of domains $\hat\xi$ over which winding number changes 
by one, and get the accumulated phase from square root of circumference of the whole 
ring of BEC pools measured in units of $\hat \xi$, as in \cite{Z}. However, the same 
result obtains from a random walk between neighboring sites, with the corresponding 
step size $\Delta\theta_s$. We now compute $\Delta\theta_s$ as a function of $\tau_Q$.

The Gross-Pitaevski equation (\ref{GPE}) can be linearized in small fluctuations 
$\delta\phi_s$ around uniform large background, $\phi_s=1+\delta\phi_s$, and 
$\delta\phi_s$ can be expanded in Bogoliubov modes as
$
\delta\phi_s=\sum_k \left( b_k u_k e^{iks}+ b_k^* v_k^* e^{-iks}\right)
$ 
with pseudomomentum $k$. For constant $J$ we have 
$
b_k(t)=b_k(0)e^{-i\omega_k t}
$ 
with 
$
\omega_k=2\sqrt{J(1-\cos k)\left[1+J(1-\cos k)\right]}
$
and stationary Bogoliubov modes
$
u_k=-{\cal N}_k\left[1+2J(1-\cos k)+\omega_k\right],~
v_k={\cal N}_k
$
where ${\cal N}_k$ are such that $u_k^2-v_k^2=1$. In the Josephson regime, when 
$J\ll1$, we have $v_k\approx -u_k$ so that purely imaginary $\delta\phi_s$ in 
$\phi_s=1+\delta\phi_s$ is a phase fluctuation. However, for our random initial 
conditions (\ref{randomphases}), this linearization is justified only for short 
wavelength modes of $\phi_s$, with $k\approx\pm\pi$, for whom the modes with longer 
wavelength are a {\it locally} uniform large background. From now on we focus on the 
short wavelength modes because they determine variance of the nearest-neighbor 
$\Delta\theta_s$. 

When $k\approx\pm\pi$ and $J\ll1$, then $\omega_k\approx2\sqrt{2J}$. Early in the 
linear quench (\ref{Jt}) this $\omega_k$ is small, so that early evolution of the 
short wavelength modes is approximately impulse i.e. their magnitude remains 
the same as in the initial Mott state and, consequently, 
$\overline{\Delta\theta_s^2}\simeq 1$ in this impulse stage. 
The impulse approximation breaks down at $\hat J$ (\cite{Z}) when the transition 
rate $\dot\omega_k/\omega_k$ equals $\omega_k$,
\be
\dot\omega_k / \omega_k ~\simeq~ \omega_k ~,
\label{hatJKZ}
\ee
and evolution becomes adiabatic. Eq. (\ref{hatJKZ}) leads to
\be
\hat J~\simeq~\tau_Q^{-2/3}~
\label{hatJ}
\ee 
which is consistent with $J\ll1$ when $\tau_Q\gg 1$. 

The crossover from impulse to adiabatic evolution at $\hat J$ is the key ingredient 
of KZM. In the following adiabatic evolution after $\hat J$ but before $J\approx1$, 
short wavelength phase fluctuations scale as $\delta\phi_s\sim J^{-1/4}$ because 
the mode amplitudes $|b_k|$ do not change, but $u_k$ and $v_k$ follow stationary 
Bogoliubov modes $u_k\approx-v_k\approx-1/2(2J)^{1/4}$. Consequently, 
$\Delta\theta_s$ has variance scaling as
$
\left.\overline{\Delta\theta_s^2}\right|_J \simeq
\overline{|\delta\phi_s|^2} \sim 
J^{-1/2}
$. 
Given the boundary condition at $\hat J$ that 
$\left.\overline{\Delta\theta_s^2}\right|_{\hat J}\simeq 1$, phase fluctuations 
must shrink as
$
\left.\overline{\Delta\theta_s^2} \right|_J                          \simeq
\left.\overline{\Delta\theta_s^2} \right|_{\hat J}~(J/\hat J)^{-1/2} \simeq
\tau_Q^{-1/3} J^{-1/2}~
$
while $J\ll 1$. 

On the other hand, when $J\gg1$ then stationary modes $u_k\approx1$ and $v_k\approx0$ 
do not depend on $J$ and $\overline{\Delta\theta_s^2}$ does not depend on $J$ either.
This means that $\overline{\Delta\theta_s^2}$ must stabilize between the regimes of 
$J\ll1$ and $J\gg1$ i.e. around $J\simeq1$ where it takes its final value  
\be
\left.\overline{\Delta\theta_s^2}\right|_{J\gg 1}     ~\approx~
\left.\overline{\Delta\theta_s^2}\right|_{J\simeq 1}  ~\simeq~
\tau_Q^{-1/3}~
\label{the 1/3}
\ee
which scales with a power of $w=1/3$.

\begin{figure}[t]
\includegraphics[width=0.99\columnwidth,clip=true]{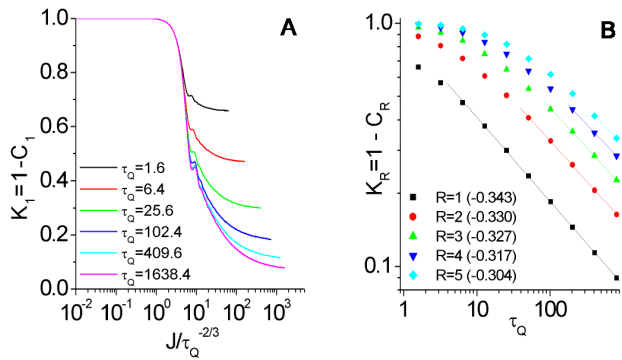}
\caption{ 
Kinetic hopping energy 
$K_1=1-C_1\approx\overline{1-\cos\Delta\theta_s}\approx\overline{\frac12\Delta\theta_s^2}$ 
as a function of rescaled $J/\hat J$ for different $\tau_Q$ is seen in A. When $J\ll 1$, all 
the plots overlap demonstrating that $\hat J=\tau_Q^{-2/3}$ is the relevant scale 
for $J\ll 1$. Individual plots depart from this small-$J$ ``common bundle'' at 
$J\simeq 1$, or $J/\tau_Q^{-2/3}\simeq \tau_Q^{2/3}$, when $K_1=1-C_1$ is expected 
to stabilize. In B, we show $K_R\equiv1-C_R$ at $J=10$ as a function of $\tau_Q$ for 
$R=1,...,5$. Data points for each $R$ were fitted with lines, their slopes giving 
exponents close to the $\frac13$-scaling predicted in Eq. (\ref{1/3-K1}) with error 
bars on their last digits. Size of typical phase step can be estimated as 
$\overline{\Delta\theta_s^2}\simeq K_R/R$ when $K_R\ll 1$, and already this rough 
estimate yields a good approximation of winding numbers shown in Fig. \ref{FigW1/3}. 
}
\label{fighatJ}
\end{figure}

This variance determines e.g. the correlator $C_1$ in 
\be
K_1~=~1-C_1=1-\overline{\cos\Delta\theta_s}~\simeq~\tau_Q^{-1/3}~,
\label{1/3-K1}
\ee
for $\tau_Q\gg 1$. Kinetic hopping energy per particle $K_1$
is expected to stabilize for $J\gg 1$, when the hopping term dominates 
over the non-linearity in Eq. (\ref{GPE}) and $K_1$ becomes an approximate 
constant of motion, see Fig. \ref{fighatJ}. 

Key ingredients of KZM are confirmed by our simulations: Phase performs a random walk 
that is markovian to a good approximation. Moreover -- as seen in Fig. \ref{fighatJ} 
-- its size is consistent with the above predictions.

\begin{figure}[t]
\includegraphics[width=1.3\columnwidth,clip=true]{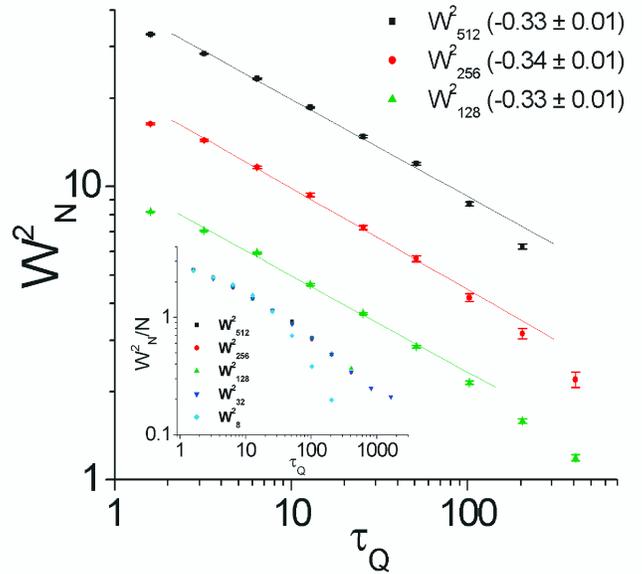}
\caption{ 
Variance of winding number $\overline{W_N^2}$ measured at $J=10$ as a function 
of $\tau_Q$ for lattice sizes $N=512,256,128$. Here point sizes equal error bars. 
The data points with $\tau_Q>2$ and $\overline{W_N^2}>2$ were fitted with the solid 
lines giving slopes close to the predicted $-\frac13$ in Eq. (\ref{W 1/3}). 
$\overline{W_N^2}$ is shown over a wider range of $\tau_Q$ to show the saturation 
for nearly instantaneous quenches, when $\tau_Q<2$, and the crossover to steeper 
slope when $\overline{W_N^2}<2$. In the inset we show a rescaled 
$\overline{W^2_N}/N$ for $N=512,256,128,32,8$ to demonstrate that 
$\overline{W^2_N}\sim N$ in the KZM regime of $\tau_Q>2$ and $\overline{W_N^2}>2$. 
}
\label{FigW1/3}
\end{figure}

{\bf Winding number. ---} Condensate wavefunction is single-valued. Therefore,
phase accumulated $\Theta_R$ after $R=N$ steps defines integer winding number:
\be
W_N~=~
\frac{1}{2\pi}
\sum_{s=1}^N {\rm Arg}\left( \phi_{s+1}\phi^*_s \right)~,
\label{WN}
\ee
where ${\rm Arg}(...)\in(-\pi,\pi]$. A random walk of phase, with the variance of 
nearest neighbor phase differences scaling as in Eq. (\ref{the 1/3}), gives winding 
numbers with variance
\be
\overline{W_N^2}~\simeq~
N~\tau_Q^{-1/3}~.
\label{W 1/3}
\ee
There are two limits where this scaling is bound to fail. For very fast quenches 
with $\tau_Q\ll 1$ phases are completely random between neighboring sites, so 
$\overline{\Delta\theta_s^2}=\pi^2/3$, and $\overline{W_N^2}=N/12$. For quenches so 
slow that $\overline{W_N^2}<1$ the nature of the problem changes, leading to 
steeper falloff of $\overline{W_N^2}$ with $\tau_Q$ \cite{lowTc,KZIsing}. Between
these two limits the $\frac13$-scaling in Eq. (\ref{W 1/3}) for the winding number 
is confirmed by our numerical results in Fig. \ref{FigW1/3}.

{\bf Correlation function. ---}
Constant amplitude and Gaussian distribution of phase $\Theta_R$ after $R$ 
steps imply
\be
C_R ~\simeq~ 
\int_{-\infty}^{\infty} 
\frac{d\Theta_R \cos\Theta_R}{\sqrt{2\pi\sigma^2_R}}~ 
e^{-\Theta_R^2/2\sigma^2_R }
~=~ 
e^{-\sigma^2_R/2}~,
\label{CRint}
\ee
where $\sigma_R$ is dispersion of 
$
\Theta_R=\sum_{s=1}^R {\rm Arg}\left( \phi_{s+1}\phi^*_s \right)
$ 
which after $R=N$ steps becomes the winding number in Eq.(\ref{WN}) i.e. $W_N=\Theta_N/2\pi$. 
For a random walk 
$\sigma^2_R=R\overline{\Delta\theta_s^2}$, which leads one to expect: 
\be
C_R ~\simeq~ 
\exp\left(-R\overline{\Delta\theta_s^2}/2\right) ~\equiv~
\exp\left(-R/\xi\right)~,
\label{CRexp}
\ee
Using Eq. (\ref{the 1/3}) we would expect scaling $\xi\simeq\tau_Q^{1/3}$. 

\begin{figure}[t]
\includegraphics[width=0.99\columnwidth,clip=true]{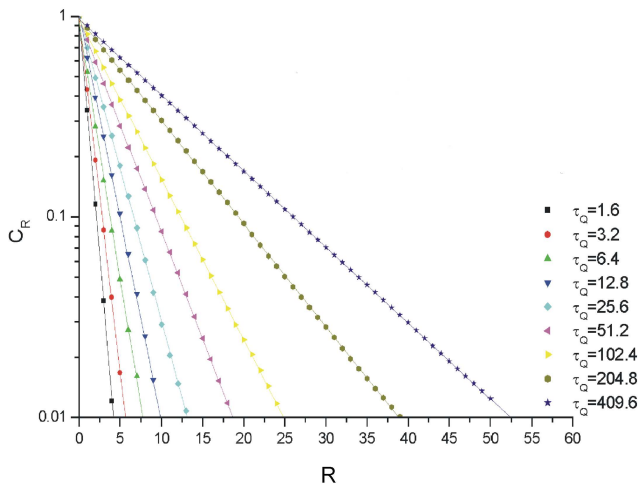}
\caption{ 
Correlation function $C_R$ for different $\tau_Q$ measured at $J=10$ on a lattice 
of $N=512$ sites. This logarithmic plot demonstrates that $C_R$ is exponential in 
$R$. 
}
\label{FigCR}
\end{figure}

Numerical simulations confirm exponential correlations, see Fig. \ref{FigCR}, 
but correlation lengths $\xi$ measured at $J=10$ are better fitted by 
$\xi\simeq\tau_Q^{0.45}$. On the other hand, early on in the quench, for smaller 
values of $J\ll 1$, correlation length exhibits $\xi\simeq\tau_Q^{1/3}$. It seems that 
intermediate scales are subject to phase ordering between the 
freezeout at $\hat J\simeq\tau_Q^{-2/3}$ and the final $J=10$. Similar post-transition 
phase ordering was observed in the integrable quantum Ising chain \cite{Cincio}. 

On the other hand, winding number continues to scale with $\tau_Q^{-1/3}$,
see Fig. \ref{FigW1/3}. It is not too surprising that it is insensitive to 
phase ordering: While in our simulations winding number is not really stable 
following the freezeout, it changes much less frequently than smaller scale 
excitations, as its topological nature leads one to expect.

{\bf Summary. ---}
We have investigated the process making a single condensate wavefunction out of many -- $N$ -- 
independent BEC pools. We conclude that, in the ring geometry, the overall winding 
number $W_N$ (which will set up persistent current) can be predicted using simple 
idea of a random walk in phase between the initially independent BEC fragments \cite{Z}. 
For very quick quenches this leads to saturation at $W_N=N/12$. Slower quenches lead 
to scaling of $W_N$ with the rate of reconnection that can be inferred from the 
Kibble-Zurek mechanism. 

Correlation functions also exhibit behavior consistent with a random walk in phase. 
Initially, correlations scale in a way that is directly related to healing length at 
the instant when dynamics of the system becomes faster than the rate of change of 
its Hamiltonian \cite{Z}. However, while winding number ``remembers'' this scaling 
as Josephson couplings increase, correlations on smaller scales evolve. In 
thermodynamic transitions similar phase ordering associated with diffusion is 
responsible for the post-transition smoothing of the order parameter structure, so 
that -- eventually -- only topological defects still ``remember'' initial state of 
the system. In our model evolution is completely reversible. Therefore, diffusion 
cannot smooth out small scale structures. However, evolution itself appears to 
redistribute energy between the excitations. This may be regarded as a quantum 
analogue of phase ordering. Correlations on intermediate scales change, but (as 
was also the case in thermodynamic phase transitions) small-scale evolution does 
not affect the topologically protected winding number $W_N$.

Our model ignores decoherence and damping that are likely to intervene in the 
laboratory experiments with, say, gaseous BECs. It is relatively easy to modify 
equations and introduce damping ``by hand''. There is however no unique prescription 
for it (although one could appeal to presence of a dilute thermal cloud, as in simple 
models of BEC decoherence \cite{DDZ}). In experiments dissipation and decoherence are 
inevitable. We expect dissipation to affect small scales, but leave the topologically 
conserved $W_N$ intact. This is based on a limited number of simulations we have 
conducted where different models of dissipation were tried out. Above all, this is 
corroborated by the experiment \cite{Anderson} where sudden reconnection of $N=3$ 
uncorrelated condensates led to relaxation to a condensate with stable vortices -- 
stable winding number. It is also consistent with the recent numerical results 
\cite{sols}.  

{\bf Acknowledgements. ---} This work was supported by DoE LDRD program at Los 
Alamos, the Polish Government project N202 079135, and 
the Marie Curie ATK project COCOS (MTKD-CT-2004-517186).

\end{document}